\newcommand{\CLASSINPUTtoptextmargin}{0.8in}
\documentclass[conference]{IEEEtran}
\usepackage{amsmath,amsfonts,amssymb,mathtools,dsfont,multicol,multirow,hhline,diagbox,dsfont,mathtools}
\usepackage{graphicx}
\usepackage{algorithmic,algorithm,threeparttable}
\usepackage{amsthm}
\usepackage[noadjust]{cite}
\newtheorem{lemma}{Lemma}
\newtheorem{prop}{Proposition}

\newcommand{\Fopt}{\mathbf{F}_\mathrm{opt}}
\newcommand{\FRF}{\mathbf{F}_\mathrm{RF}}

\newcommand{\FBB}{\mathbf{F}_\mathrm{BB}}
\newcommand{\FDD}{\mathbf{F}_\mathrm{DD}}

\newcommand{\WBB}{\mathbf{W}_\mathrm{BB}}
\newcommand{\WRF}{\mathbf{W}_\mathrm{RF}}
\newcommand{\NRFr}{N_\mathrm{RF}^\mathrm{r}}
\newcommand{\NRFt}{N_\mathrm{RF}^\mathrm{t}}
\newcommand{\Nt}{N_\mathrm{t}}
\newcommand{\Nr}{N_\mathrm{r}}
\DeclareMathOperator{\Tr}{tr}

\newcommand\relphantom[1]{\mathrel{\phantom{#1}}}

\newcounter{longequ}[longequ]
\addtocounter{longequ}{25}

\ifCLASSINFOpdf
\else
\fi

\IEEEoverridecommandlockouts

\begin{document}
	\title{Hybrid Precoding in Millimeter Wave Systems: How Many Phase Shifters Are Needed?}
	\author{\IEEEauthorblockN{Xianghao Yu$^*$, Jun Zhang$^*$, and Khaled B. Letaief$^{*\dag}$, \emph{Fellow, IEEE} }
		\IEEEauthorblockA{$^*$Dept. of ECE, The Hong Kong University of Science and Technology, Hong Kong\\
			$^\dag$Hamad Bin Khalifa University, Doha, Qatar\\
			Email: $^*$\{xyuam, eejzhang, eekhaled\}@ust.hk, $^\dag$kletaief@hbku.edu.qa}
		\thanks{This work was supported by the Hong Kong Research Grants Council under Grant No. 16210216. 
		}
	}
	
	\maketitle
	
	\begin{abstract}
		Hybrid precoding has been recently proposed as a cost-effective transceiver solution for millimeter wave (mm-wave) systems. 
		The analog component in such precoders, which is composed of a phase shifter network, is the key differentiating element in contrast to conventional fully digital precoders. 
		While a large number of phase shifters with unquantized phases are commonly assumed in existing works, in practice the phase shifters should be discretized with a coarse quantization, and their number should be reduced to a minimum due to cost and power consideration.
		In this paper, we propose a new hybrid precoder implementation using a small number of phase shifters with quantized and \emph{fixed} phases, i.e., a \emph{fixed phase shifter} (FPS) implementation, which significantly reduces the cost and hardware complexity.
		In addition, a dynamic switch network is proposed to enhance the spectral efficiency. 
		Based on the proposed FPS implementation, an effective alternating minimization (AltMin) algorithm is developed with closed-form solutions in each iteration. Simulation results show that the proposed algorithm with the FPS implementation outperforms existing ones. More importantly, it needs much fewer phase shifters than existing hybrid precoder proposals, e.g., $\sim$10 fixed phase shifters are sufficient for practically relevant system settings. 
	\end{abstract}
	
	\IEEEpeerreviewmaketitle
	
	\section{Introduction}
	Uplifting the carrier frequency to millimeter wave (mm-wave) bands has been proposed to meet the capacity requirement of the upcoming 5G networks, and it thus has drawn extensive attention from both academia and industry \cite{6515173}. Thanks to the small wavelength of mm-wave signals, large-scale antenna arrays can be leveraged at transceivers to support directional transmissions. As equipping each antenna element with a single radio frequency (RF) chain is costly, hybrid precoding has been put forward as a cost-effective solution, which utilizes a limited number of RF chains to incorporate a digital baseband precoder and an analog RF precoder \cite{6717211}.
	
	In contrast to the conventional fully digital precoder, the additional component in the hybrid one is the analog precoder, which is usually implemented by a bunch of phase shifters in the RF domain.
	%and serves as the beamforming gain for each connected link from RF chains to antennas. forms the main challenge in the hybrid precoder design. 
	Several hybrid precoder structures and implementations have been proposed in existing works, e.g., the fully- and partially-connected structures \cite{7397861}, as well as the single phase shifter (SPS) \cite{6717211} and double phase shifter (DPS) \cite{asilomar} implementations, to provide trade-off between spectral efficiency, energy efficiency, and algorithmic complexity. The main differences among them are the connecting strategies from RF chains to antennas and the number of phase shifters in use to compose the beamforming gain for each of the connected paths. While existing hybrid precoder structures and implementations enjoy a small number of RF chains, the number of phase shifters scales linearly with the antenna size, which is a huge number and thus causes prohibitively high cost and power consumption.
	On the other hand, various hybrid precoding algorithms have been proposed assuming phase shifters with arbitrary precision, e.g., orthogonal matching pursuit (OMP) \cite{6717211}, manifold optimization \cite{7397861}, and successive interference canceling \cite{7445130}.
	Although considering phase shifters with programmable high resolution eases the hybrid precoder design, it will weaken the practicality of the results since adaptively carrying out arbitrary phase shifts at mm-wave frequencies is highly impractical \cite{5648370}. Therefore, it is of critical importance to develop effective design methodologies for hybrid precoders with a \emph{small number} of \emph{quantized} phase shifters.
	
	There are a few works that attempted to consider quantized phase shifters \cite{6717211,7827111,6928432,7370753,7387790,7858800}. 
	The main approach is either to determine all the phases at once \cite{6717211,7827111,6928432,7370753,7387790} or update one phase at a time \cite{7858800} by ignoring the quantization effect at first. Then the phases are heuristically quantized into the finite feasible set according to a certain criterion. However, a simple quantization step is far from satisfactory, and the optimality and convergence of the proposed algorithms cannot be guaranteed \cite{7858800}. On the other hand, the number of phase shifters in use was to some extent reduced in \cite{7387790}, which was determined for achieving a certain precision of the unquantized ones. Unfortunately, a large number of phase shifters are still needed for practical settings, i.e., 40 phase shifters for each RF chain, and the number will vary according to the precision requirement.
	More importantly, in existing works, the phases need to be adapted to the channel states, which brings high complexity for hardware implementation and also increases power consumption.
	%More importantly, although the quantization effects are considered in existing works, there still exist two main issues that need to be resolved: 1) the number of phase shifters scales with the antenna size, which is a huge number and thus causes prohibitively high cost and power consumption; 2) The phase shifters need to be adaptive to the channel states, which places high demands on transceiver hardware design and implementation.
	\begin{figure*}
		\centering
		\includegraphics[height=4.5cm]{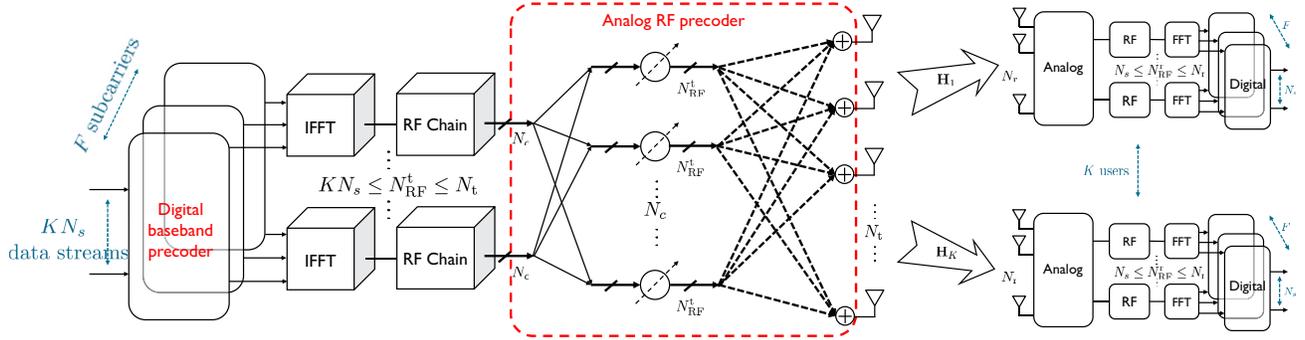}
		\caption{A multiuser mm-wave MIMO-OFDM system with FPS hybrid precoder implementation. To simplify the figure, in the analog precoder, each solid line with a slash represents parallel signals transmission while each dotted line stands for $\NRFt$ switches. 
			%		The structures of the combiners are similar to that of the precoder, and is omitted due to space limitation.
		}
		\label{systemmodel}
	\end{figure*}
	
	To overcome the above limitations, in this paper we propose a novel hybrid precoder implementation for general multiuser orthogonal frequency-division multiplexing (OFDM) mm-wave systems, where only a small number of phase shifters with \emph{fixed} phases are available \cite{5648370}, namely the fixed phase shifter (FPS) implementation. To compensate the performance loss induced by the fixed phases, a switch network is proposed to provide dynamic mappings from phase shifters to antennas, which is easily implementable with adaptive switches \cite{5648370,7370753}. With the proposed FPS implementation, we develop an alternating minimization (AltMin) algorithm to design the hybrid precoder \cite{7397861}, where an upper bound of the objective function is derived as an effective surrogate. In particular, the large-scale binary constraints introduced by the switch network are delicately tackled with the help of the upper bound, which leads to closed-from solutions for both the dynamic switch network and the digital baseband precoder, and therefore enables a low-complexity hybrid precoding algorithm. Simulation results shall demonstrate that the proposed FPS-AltMin algorithm outperforms existing ones and approaches the performance of the fully digital precoder. What deserves a special mention is the sharp reduction of the number of phase shifters compared to existing hybrid precoder implementations, which indicates that the proposed FPS implementation is a promising candidate for hybrid precoding in 5G mm-wave communication systems.
	
	\section{System Model}
	\subsection{Signal Model}
	Consider the downlink transmission for a multiuser mm-wave MIMO-OFDM system as shown in Fig. \ref{systemmodel}. A base station (BS) leverages an $\Nt$-size antenna array to serve $K$ users over $F$ subcarriers using OFDM. Each user is equipped with $\Nr$ antennas and receives $N_s$ data streams from the BS on each subcarrier. The numbers of available RF chains are $\NRFt$ and $\NRFr$ for the BS and each user, respectively, which are restricted as $KN_s\le\NRFt<\Nt$ and $N_s\le\NRFr<\Nr$.
	
	The received signal of the $k$-th user on the $f$-th subcarrier is given by
	\begin{equation}
	\mathbf{y}_{k,f}=\mathbf{W}^H_{\mathrm{BB}k,f}\mathbf{W}^H_{\mathrm{RF}k}\left(\mathbf{H}_{k,f}\FRF\sum_{k=1}^K{\FBB}_{k,f}\mathbf{s}_{k,f}+\mathbf{n}_{k,f}\right),
	\end{equation}
	where $\mathbf{s}_{k,f}$ is the transmitted signal to the $k$-th user on the $f$-th subcarrier such that $\mathbb{E}\left[\mathbf{s}_{k,f}\mathbf{s}_{k,f}^H\right]=\frac{P}{KN_sF}\mathbf{I}_{N_s}$, and $\mathbf{n}_{k,f}$ is the circularly symmetric complex Gaussian noise with power as $\sigma_\mathrm{n}^2$ at the users. The digital baseband precoders and combiners are denoted as ${\FBB}_{k,f}$ and ${\WBB}_{k,f}$, respectively, with dimensions $\NRFt\times N_s$ and $\NRFr\times N_s$. Since the transmitted signals for all the users are mixed together by the digital precoders, and analog RF precoding is a post-IFFT operation, the RF analog precoder $\FRF$ with dimension $\Nt\times\NRFt$ is a common component for all the users and subcarriers. Correspondingly, the $\Nr\times\NRFr$ RF analog combiner ${\WRF}_k$ is subcarrier-independent for each user.
	
	\subsection{FPS Implementation}
	In earlier works on hybrid precoding \cite{6717211,7397861,7445130,7827111,6928432,7370753}, a single phase shifter is adopted to adjust the phase of each of the paths from RF chains to antennas. Therefore, $\Nt\NRFt$ phase shifters are required, commonly assumed with arbitrary precision. 
	Recently, it was shown in \cite{asilomar} that the performance of the hybrid precoder can be greatly improved by passing each signal through two unquantized phase shifters and then combining the outputs, which, however, induces high hardware complexity by employing $2\Nt\NRFt$ adaptive phase shifters. 
	
In this paper, we propose a hybrid precoder implementation using $N_c$ phase shifters with \emph{fixed} phases \cite{5648370}, where $N_c\ll\Nt\NRFt$, as shown in Fig. \ref{systemmodel}. 
	Nevertheless, the limited number of fixed phase shifters, which cannot be adaptively adjusted according to the channel states, inevitably entail performance loss. To overcome this drawback brought by the simplified hardware implementation, we propose to cascade a dynamic switch network after the fixed phase shifters, which is adapted to the channel states.
	
\begin{figure}[t]
		\centering
		\includegraphics[height=3cm]{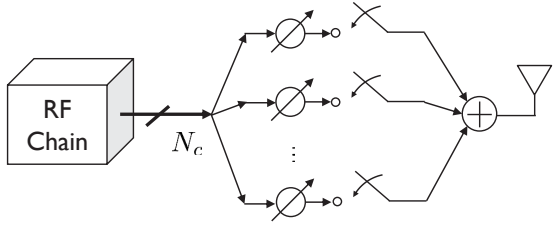}
		\caption{The FPS implementation from an RF chain to each antenna.
		}
		\label{sub}
	\end{figure}

%In particular, in each of the fixed phase shifters, the $\NRFt$ distinct signals from different RF chains are processed one by one, i.e., in a time division multiplexing (TDM) fashion. 
%In particular, in the phase shifter network, each of the output signals from $\NRFt$ RF chains is simultaneously processed by $N_c$ $\NRFt$-channel fixed phase shifters \cite{6690170}, i.e., in a parallel fashion.
In particular, $N_c$ multichannel ($\NRFt$-channel) fixed phase shifters \cite{6690170} are deployed in the phase shifter network, each of which simultaneously processes the output signals from $\NRFt$ RF chains, i.e., in a parallel fashion.
To clearly illustrate the proposed FPS implementation, we focus on one signal flow from an RF chain to an antenna, as shown in Fig. \ref{sub}.
The $N_c$ fixed phase shifters generate $N_c$ signals with different phases for the output signal of the given RF chain. Inspired by the idea of doubling phase shifters to achieve high spectral efficiency, as demonstrated in \cite{asilomar}, we propose to adaptively combine a subset of the $N_c$ signals to compose the analog precoding gain from the RF chain to the antenna, which is implemented with $N_c$ adaptive switches. As $N_c$ switches are needed for each RF chain-antenna pair, in total $\Nt \NRFt N_c$ switches are required in the proposed FPS implementation. Note that adaptive switches with binary states are easier to implement in mm-wave bands than adaptive phase shifters with arbitrary precision \cite{5648370,7370753}. 
	
	Accordingly, the analog RF precoding matrix $\FRF$ can be expressed as
	\begin{equation}
	\FRF=\mathbf{SC},
	\end{equation}
	where $\mathbf{S}\in\{0,1\}^{\Nt\times N_c\NRFt}$ is the switch matrix, and the boolean constraints are induced by the switches with binary states. The matrix $\mathbf{C}\in\mathbb{C}^{N_c\NRFt\times\NRFt}$ stands for the phase shift operation carried out by the available fixed phase shifters, given by a block diagonal matrix as
	\begin{equation}\label{psmatrix}
	\mathbf{C}=\mathrm{diag}\left(\underbrace{\mathbf{c},\mathbf{c},\cdots,\mathbf{c}}_{\NRFt}\right),
	\end{equation}
	where $\mathbf{c}=\frac{1}{\sqrt{N_c}}\left[e^{\jmath\theta_1},e^{\jmath\theta_2},\cdots,e^{\jmath\theta_{N_c}}\right]^T$ is the normalized phase shifter vector containing all $N_c$ fixed phases $\left\{\theta_i\right\}_{i=1}^{N_c}$.
	
	\subsection{Problem Formulation}
	It has been shown in \cite{6717211,7397861,7827111,7858800,asilomar} that minimizing the Euclidean distance between the fully digital precoder and the hybrid precoder is an effective and tractable alternative objective for maximizing the spectral efficiency of mm-wave systems. In this paper, we resort to this approach and the hybrid precoder design is correspondingly formulated as
	\begin{equation}
	\mathcal{P}_1:\quad
	\begin{aligned}
	&\underset{\mathbf{S},\mathbf{F}_\mathrm{BB}}{\mathrm{minimize}} && \left\Vert\Fopt-\mathbf{SC}\FBB\right\Vert_F^2\\
	&\mathrm{subject\thinspace to}&&
	\begin{cases}
	\mathbf{S}\in\mathcal{B}\\
	\left\Vert\mathbf{SC}\FBB\right\Vert_F^2\le KN_sF,\\
	\end{cases}
	\end{aligned}
	\end{equation}
	where $\Fopt=\left[{\Fopt}_{1,1},\cdots,{\Fopt}_{k,f},\cdots,{\Fopt}_{K,F}\right]$ is the combined fully digital precoder with dimension ${\Nt\times KN_sF}$, and $\FBB=\left[{\FBB}_{1,1},\cdots,{\FBB}_{k,f},\cdots,{\FBB}_{K,F}\right]$ is the concatenated digital baseband precoder with dimension ${\NRFt\times KN_sF}$. The set of binary matrices is denoted as $\mathcal{B}$, and the second constraint is the transmit power constraint. Note that the combiners at the user side can be designed in the same way without the power constraint \cite{7397861,7389996} and thus are omitted due to space limitation.
	
	\emph{Remark 1:} Since the switch matrix $\mathbf{S}$ is with finite possibilities, the cardinality of the constraint set for $\FRF$ is finite, which means the OMP algorithm \cite{6717211} is applicable to this problem $\mathcal{P}_1$. However, the dimension of the dictionary in the OMP algorithm is oversize, i.e., $\left[\sum_{i=1}^{N_c}\binom{N_c}{i}\right]^{\Nt}$, which prevents its practical implementation.
	
	\emph{Remark 2:} Alternating minimization can be directly applied to $\mathcal{P}_1$ where the binary constraints can be tackled with semidefinite relaxation \cite{7397861}. However, an $\Nt\NRFt N_c+1$-dimension semidefinite programming (SDP) problem should be solved in each iteration, which causes prohibitive computational complexity. Moreover, the optimality of the relaxation in each iteration cannot be ensured and hence the overall convergence of the AltMin algorithm cannot be guaranteed.
	
	As illustrated above, the main difficulty to solve $\mathcal{P}_1$ is the binary constraints of $\mathbf{S}$, and it is the main obstacle for designing an efficient algorithm with performance guarantee. In this paper, by deriving an effective surrogate and adopting alternating minimization, we shall come up with a low-complexity hybrid precoding algorithm that well tackles the binary constraints.
	
	\section{Hybrid Precoder Design in Single-Carrier Systems With the FPS Implementation}
	In this section, we first present the hybrid precoder design in single-carrier systems\footnote{In this paper, single-carrier systems refer to single-carrier transmissions assuming flat-fading channels. The choice of such systems is for the ease of presentation, and the algorithm will be later extended to the more realistic multicarrier case with frequency-selective fading channels.}, i.e., when $F=1$. In particular, an upper bound of the objective function is firstly derived, based on which an alternating minimization algorithm is then developed.
	\subsection{Objective Upper Bound}
	In \cite{7397861,asilomar,7389996}, imposing a semi-orthogonal structure for $\FBB$ is shown to achieve near-optimal performance. Inspired by these results, we take a similar approach. In single-carrier systems, the digital precoder matrix $\FBB$ is with dimension $\NRFt\times KN_s$. Recall that the number of RF chains is limited as $KN_s\le\NRFt<\Nt$, which forces $\FBB$ as to be tall matrix, and thus the semi-orthogonal constraint is specified as
	\begin{equation}\label{eq5}
	\FBB^H\FBB=\alpha^2\FDD^H\FDD=\alpha^2\mathbf{I}_{KN_s},
	\end{equation}
	where $\FBB=\alpha\FDD$ and $\FDD$ is a semi-unitary matrix. Then, the objective function in $\mathcal{P}_1$ can be rewritten as
	\begin{equation}\label{obj}
	\left\Vert\Fopt\right\Vert_F^2-2\alpha\Re\Tr\left(\FDD\Fopt^H\mathbf{SC}\right)+\alpha^2\left\Vert\mathbf{SC}\FDD\right\Vert_F^2.
	\end{equation}
	
	Note that, according to \eqref{psmatrix}, the phase shifter matrix $\mathbf{C}$ is also a semi-unitary matrix, i.e., $\mathbf{C}^H\mathbf{C}=\mathbf{I}_{\NRFt}$. Therefore, we can derive an upper bound for the last term in \eqref{obj}, given by
	\begin{equation}\label{upper}
	\begin{split}
	\left\Vert\mathbf{SC}\FDD\right\Vert_F^2&=
	\Tr\left(\FDD^H\mathbf{C}^H\mathbf{S}^H\mathbf{SC}\FDD\right)\\
	&\overset{(a)}{=}\Tr \left( {
		\begin{bmatrix}
		\mathbf{I}_{KN_s}&\\
		&\mathbf{0}\\
		\end{bmatrix}
		{{\mathbf{K}^H}}{{\mathbf{S}^H}}{\mathbf{S}\mathbf{K}}} \right)\\
	&< \Tr\left( {{{\mathbf{K}}^H}{\mathbf{S}^H}}{\mathbf{S}\mathbf{K}} \right)=\left\Vert\mathbf{S}\right\Vert_F^2,
	\end{split}
	\end{equation}
	where (a) follows the singular value decomposition (SVD) of $\mathbf{C}\FDD\FDD^H\mathbf{C}^H=\mathbf{K}\mathrm{diag}\left(\mathbf{I}_{KN_s},\mathbf{0}\right)\mathbf{K}^H
	%=\mathbf{K}\begin{bmatrix}\mathbf{I}_{KN_s}&\\&\mathbf{0}\\\end{bmatrix}\mathbf{K}^H
	$ by utilizing the semi-unitary property of $\mathbf{C}\FDD$. Thus, we obtain an upper bound for the original objective function, expressed as
	\begin{equation}\label{upperobj}
	\left\Vert\Fopt\right\Vert_F^2-2\alpha\Re\Tr\left(\FDD\Fopt^H\mathbf{SC}\right)+\alpha^2\left\Vert\mathbf{S}\right\Vert_F^2.
	\end{equation}
	%In the following, we will see the benefits of adopting this upper bound as the surrogate objective function.
	
	\subsection{Alternating Minimization}
	By adopting the upper bound \eqref{upperobj} as the surrogate objective function and dropping the constant term $\left\Vert\Fopt\right\Vert_F^2$, the hybrid precoder design problem is reformulated as
	\begin{equation}
	\mathcal{P}_2:\quad
	\begin{aligned}
	&\underset{\alpha,\mathbf{S},\mathbf{F}_\mathrm{DD}}{\mathrm{minimize}} && \alpha^2\left\Vert\mathbf{S}\right\Vert_F^2-2\alpha\Re\Tr\left(\FDD\Fopt^H\mathbf{SC}\right)\\
	&\mathrm{subject\thinspace to}&&
	\begin{cases}
	\mathbf{S}\in\mathcal{B}\\
	\FDD^H\FDD=\mathbf{I}_{KN_s}.
	\end{cases}
	\end{aligned}
	\end{equation}
	%Note that the power constraint in $\mathcal{P}_1$ is temporarily removed. In fact, a simple normalization operation can be adopted if the power constraint is not satisfied, which will be discussed later.
	
	Alternating minimization, as an effective tool for optimization problems involving different subsets of variables, has been widely applied and shown empirically successful in hybrid precoder design \cite{7397861,asilomar,7389996}. In this section, we apply it this effective rule of thumb to the hybrid precoder design with the FPS implementation.
	
	In each step of the AltMin algorithm, one subset of the optimization variables is optimized while keeping the other parts fixed. When the switch matrix $\mathbf{S}$ and $\alpha$ are fixed, the optimization problem can be written as
	\begin{equation}
	\begin{aligned}
	&\underset{\mathbf{F}_\mathrm{DD}}{\mathrm{maximize}} && \alpha\Re\Tr\left(\FDD\Fopt^H\mathbf{SC}\right)\\
	&\mathrm{subject\thinspace to}&&
	\FDD^H\FDD=\mathbf{I}_{KN_s}.
	\end{aligned}
	\end{equation}
	According to the definition of the dual norm \cite{horn2012matrix}, we have
	\begin{equation}
	\begin{split}
	\alpha\Re\Tr\left(\FDD\Fopt^H\mathbf{SC}\right)&\le\left|\Tr\left(\alpha\FDD\Fopt^H\mathbf{SC}\right)\right|\\
	&\overset{(b)}{\le}\left\Vert\FDD^H\right\Vert_\infty\left\Vert\alpha\Fopt^H\mathbf{SC}\right\Vert_1\\
	&=\left\Vert\alpha\Fopt^H\mathbf{SC}\right\Vert_1=\sum_{i=1}^{KN_s}{\sigma_i},
	\end{split}
	\end{equation}
	where $\left\Vert\cdot\right\Vert_\infty$ and $\left\Vert\cdot\right\Vert_1$ stand for the infinite and one Schatten norms \cite{horn2012matrix}, and (b) follows the H\"{o}lder's inequality. The equality is established only when
	\begin{equation}\label{uv}
	\FDD=\mathbf{V}_1\mathbf{U}^H,
	\end{equation}
	where $\alpha\Fopt^H\mathbf{SC}=\mathbf{U}\mathbf{\Sigma V}_1^H$ follows the SVD and $\mathbf{\Sigma}$ is a diagonal matrix with non-zero singular values $\sigma_1,\cdots,\sigma_{KN_s}$.
	
	While we can divide the optimization of the two variables $\alpha$ and $\mathbf{S}$ into two separate subproblems, we propose to update them in parallel to save the number of subproblems involved in the AltMin algorithm and therefore reduce the computational complexity. By adding a constant term $\left\Vert\Re\left(\Fopt\FDD^H\mathbf{C}^H\right)\right\Vert_F^2$ to the objective function of $\mathcal{P}_2$, the subproblem of updating $\alpha$ and $\mathbf{S}$ can be recast as
	\begin{equation}\label{eq13}
	\begin{aligned}
	&\underset{\alpha,\mathbf{S}}{\mathrm{minimize}} && \left\Vert\Re\left(\Fopt\FDD^H\mathbf{C}^H\right)-\alpha\mathbf{S}\right\Vert_F^2\\
	&\mathrm{subject\thinspace to}&&
	\mathbf{S}\in\mathcal{B}.
	\end{aligned}
	\end{equation}
	\begin{prop}
		The optimal solution to \eqref{eq13} is given by
		\begin{equation}
		\alpha^\star=\arg\underset{\{\tilde{x}_i,{\bar{x}}_i\}_{i=1}^n}{\min}\quad\left\{f(\tilde{x}_i),f({\bar{x}}_i)\right\},
		\end{equation}
		\begin{equation}\label{eq15}
		\mathbf{S}^\star=\begin{cases}
		\mathds{1}\left\{\Re\left(\Fopt\FDD^H\mathbf{C}^H\right)>\frac{\alpha}{2}\mathbf{1}_{\Nt\times N_c\NRFt}\right\}&\alpha>0\\
		\mathds{1}\left\{\Re\left(\Fopt\FDD^H\mathbf{C}^H\right)<\frac{\alpha}{2}\mathbf{1}_{\Nt\times N_c\NRFt}\right\}&\alpha<0,\\
		\end{cases}
		\end{equation}
		where $\mathbf{x}=\mathrm{vec}\left\{\Re\left(\Fopt\FDD^H\mathbf{C}^H\right)\right\}$, $\mathds{1}(\cdot)$ is the indicator function, and $\mathbf{1}_{m\times n}$ denotes an $m\times n$ matrix with all entries equal to one. The objective function in \eqref{eq13} can be rewritten as $f(\alpha)$ as \eqref{longeq} in the proof. In addition, $\tilde{x}_i$ is the $i$-th smallest entry in $\mathbf{x}$, and\footnote{$f(\alpha)$ is a coercive function, i.e., $f(+\infty)\to+\infty$.}
		\begin{equation}\label{eq16}
		{\bar{x}}_i\triangleq\begin{cases}
		\frac{\sum_{j=1}^i\tilde{x}_j}{i}&\alpha<0\text{ and }\frac{\sum_{j=1}^i\tilde{x}_j}{i}\in[2\tilde{x}_i,2\tilde{x}_{i+1}]\\
		\frac{\sum_{j=i+1}^n\tilde{x}_j}{n-i}&\alpha>0\text{ and }\frac{\sum_{j=i+1}^n\tilde{x}_j}{n-i}\in[2\tilde{x}_i,2\tilde{x}_{i+1}]\\
		+\infty&\text{otherwiese}.
		\end{cases}
		\end{equation}
	\end{prop}
	\begin{IEEEproof}
		See Appendix A.
	\end{IEEEproof}
	Basically, the optimal $\alpha^\star$ is obtained via a closed-form solution by comparing the optimal solutions of $\alpha$ in all the intervals $\{\mathcal{R}_i\}_{i=1}^n$, where $\mathcal{R}_i\triangleq[2\tilde{x}_i,2\tilde{x}_{i+1}]$.
	Nevertheless, since the number of intervals that need to compare is $n=\Nt N_c\NRFt$, it will incur high computational complexity when $\Nt$ is large in mm-wave systems. In the following lemma, we show that there is no need to compute the optimal $\alpha$ in all the intervals $\mathcal{R}_i$, which will further reduce the complexity of the proposed algorithm.
	\begin{lemma}\label{lem1}
		The optimal $\alpha^\star$ is obtained at one of the points $\{{\bar{x}}_i\}_{i=1}^n$.
	\end{lemma}
	\begin{IEEEproof}
		See Appendix B.
	\end{IEEEproof}
	Lemma \ref{lem1} indicates that any endpoint of the intervals cannot be the optimal solution for $\alpha$. Therefore, we only need to pick the ${\bar{x}}_i$'s that have finite values of $f({\bar{x}}_i)$, i.e., the ones that satisfy the first two conditions in \eqref{eq16}, denoted as a set $\mathcal{X}$, and the optimal solution for $\alpha$ is given by
	\begin{equation}\label{eq20}
	\alpha^\star=\arg\underset{{{\bar{x}}_i\in\mathcal{X}}}{\min}\quad f({\bar{x}}_i).
	\end{equation}
	By Lemma \ref{lem1}, the number of intervals we need to compare to obtain the optimal $\alpha^\star$ is shrunk from $n$ to $|\mathcal{X}|$, which is empirically shown to be less than 5 via simulations in Section V and hence further reduces the computational complexity of the proposed AltMin algorithm. 
	
	\emph{Remark 3:}
	It is shown that, with the help of the upper bound derived in \eqref{upper}, the large-scale binary switch matrix $\mathbf{S}$ can be efficiently optimized by a closed-form solution, which verifies the benefits and superiority of the surrogate objective function adopted in $\mathcal{P}_2$.
	
	With the closed-form solutions derived in \eqref{uv}, \eqref{eq15}, and \eqref{eq20} at hands, the AltMin algorithm for the FPS implementation is summarized as \textbf{FPS-AltMin Algorithm}.
	The FPS-AltMin algorithm is essentially a block coordinate descent (BCD) algorithm with two blocks that have globally optimal solutions in Steps 3 and 4, and the algorithm is guaranteed to converge to a stationary point of $\mathcal{P}_2$ \cite{grippo2000convergence}. The algorithm may be sensitive to the initial point $\FDD^{(0)}$. 
	Note that the fully digital precoding matrix $\Fopt$ can be decomposed as follows according to its SVD $\Fopt=\mathbf{U\Sigma V}^H$, i.e,
	\begin{equation}\label{eq18}
	\Fopt=\begin{bmatrix}
	\mathbf{U\Sigma}&\mathbf{F}
	\end{bmatrix}
	\begin{bmatrix}
	\mathbf{V}^H\\
	\mathbf{0}
	\end{bmatrix},
	\end{equation} 
	where $\mathbf{U\Sigma}$ is an $\Nt\times KN_s$ full rank matrix, $\mathbf{V}^H$ is a $KN_s$ dimension square matrix, and $\mathbf{F}$ is an arbitrary $\Nt\times(\NRFt-KN_s)$ matrix. In \eqref{eq18}, the fully digital precoding matrix $\Fopt$ is decomposed into two matrices that satisfy the dimensions of $\FRF$ and $\FDD$, respectively. In this way, we propose to construct the initial point $\FDD^{(0)}$ as
	\begin{equation}
	\FDD^{(0)}=
	\begin{bmatrix}
	\mathbf{V}&\mathbf{0}_{KN_s\times(\NRFt-KN_s)}
	\end{bmatrix}^H.
	\end{equation}
	To cancel the inter-user interference, similar to \cite{asilomar}, we cascade an additional block diagonal precoder at the baseband in the Step 7 based on the effective channel including the hybrid precoder and physical channel. In the final step, we normalize the digital precoder to maximize the signal to noise ratio (SNR) while satisfying the transmit power constraint.
	\floatname{algorithm}{FPS-AltMin Algorithm:}
	\begin{algorithm}[t]
		\caption{A Low-Complexity Hybrid Precoding Algorithm for the FPS Implementation}
		\begin{algorithmic}[1]\label{lowcom}
			\REQUIRE
			$\Fopt$
			\STATE Construct an initial point for $\FDD^{(0)}$ and set $k=0$;
			\REPEAT 
			\STATE Fix $\FDD^{(k)}$, optimize $\alpha^{(k)}$ and $\mathbf{S}^{(k)}$ according to \eqref{eq20} and \eqref{eq15}, respectively;
			\STATE Fix $\mathbf{S}^{(k)}$ and $\alpha^{(k)}$, update $\FDD^{(k)}$ with \eqref{uv}; 
			\STATE $k\leftarrow k+1$;
			\UNTIL convergence.
			\STATE Compute the additional BD precoder at the baseband to cancel the inter-user interference \cite{asilomar}.
			\STATE For the digital precoder at the transmit end, normalize
			${\mathbf{F}}_\mathrm{BB}=\frac{\sqrt{KN_sF}}{\left\Vert\mathbf{SC}\FDD\right\Vert_F}\FDD.$
		\end{algorithmic}
	\end{algorithm}
	
	\section{Hybrid Precoder Design in Multicarrier Systems With the FPS Implementation}
	Multicarrier techniques such as OFDM are often utilized to overcome the multipath fading caused by the large available bandwidth in mm-wave systems. Compared with the narrowband hybrid precoder design in Section III, the main difference in OFDM systems is that the analog precoder is shared by all the subcarriers. In particular, the digital precoding matrix $\FBB\in\mathbb{C}^{\NRFt\times KN_sF}$ in $\mathcal{P}_1$ is no longer a tall matrix since $KN_sF\ge \NRFt$ for practical OFDM system settings.
	
	In this section, we modify the FPS-AltMin algorithm for OFDM systems. Similar to \eqref{eq5}, we enforce a semi-orthogonal constraint on the digital precoding matrix, i.e.,
	\begin{equation}
	\FBB\FBB^H=\alpha ^2\FDD\FDD^H=\alpha^2\mathbf{I}_{\NRFt}.
	\end{equation}
	In this way, the upper bound of the objective function derived in \eqref{upper} still holds since
	\begin{equation}
	\begin{split}
	\left\Vert\mathbf{SC}\FDD\right\Vert_F^2&=
	\Tr\left(\mathbf{C}^H\mathbf{S}^H\mathbf{SC}\right)\\
	&\overset{(c)}{=}\Tr\left( { {
			\begin{bmatrix}
			\mathbf{I}_{\NRFt}&\\
			&\mathbf{0}
			\end{bmatrix}	
		} {{\mathbf{K}^H}}{{\mathbf{S}^H}}{\mathbf{S}\mathbf{K}}} \right)\\
	&< \Tr\left( {{{\mathbf{K}}^H}{\mathbf{S}^H}}{\mathbf{S}\mathbf{K}} \right)=\left\Vert\mathbf{S}\right\Vert_F^2,
	\end{split}
	\end{equation}
	where (c) comes from the SVD of $\mathbf{CC}^H$ since $\mathbf{C}$ is a semi-unitary matrix. In the AltMin algorithm, the update of $\alpha$ and $\mathbf{S}$ is the same as that in Section III-B. Since the dimension of $\FDD$ is different in OFDM systems, the optimization of $\FDD$ is modified as
	\begin{equation}\label{eq24}
	\FDD=\mathbf{V}\mathbf{U}_1^H,
	\end{equation}
	where $\Fopt^H\mathbf{SC}=\mathbf{U}_1\mathbf{\Sigma V}^H$ and $\mathbf{\Sigma}$ is a diagonal matrix with non-zero singular values $\sigma_1,\cdots,\sigma_{\NRFt}$, which is the SVD of $\Fopt^H\mathbf{SC}$. Correspondingly, the construction of the initial $\FDD^{(0)}$ is given by
	\begin{equation}\label{eq25}
	\FDD^{(0)}=\mathbf{V}^H_{[1:\NRFt]},
	\end{equation}
	where $\Fopt=\mathbf{U}\mathbf{\Sigma V}^H$ is the SVD of $\Fopt$ and the subscript $[1:n]$ denotes the first to the $n$-th columns of a matrix.
	
	By substituting \eqref{eq25} and \eqref{eq24} into the Steps 1 and 4, we obtain the FPS-AltMin algorithm for OFDM mm-wave systems.
	
	\section{Simulation Results}
	In this section, we will evaluate the performance of the proposed FPS-AltMin algorithm via simulations. The BS and each user are equipped with 144 and 16 antennas, respectively, while all the transceivers are equipped with uniform planar arrays (UPAs). Four users and 128 subcarriers are assumed when considering multiuser OFDM systems. To reduce the cost and power consumption, the minimum number of RF chains is adopted according to the assumptions in Section II-A, i.e., $\NRFt=KN_s$ and $\NRFr=N_s$. The phases of the available fixed phase shifters are uniformly separated within $[0,2\pi]$ by $N_c$ equal length intervals. Furthermore, the Saleh-Valenzuela model is adopted in simulations to characterize mm-wave channels \cite{6717211,7397861}. The nominal SNR is defined as $\frac{P}{KN_sF\sigma_\mathrm{n}^2}$, and all the simulation results are averaged over 1000 channel realizations.
	
	\subsection{Single-User Single-Carrier (SU-SC) Systems}
	\begin{figure}
		\centering
		\includegraphics[height=5.5cm]{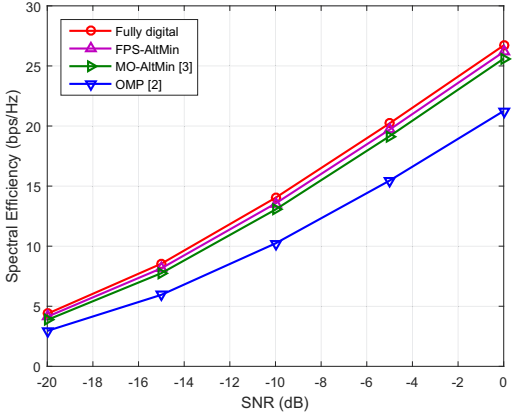}
		\caption{Spectral efficiency achieved by different hybrid precoding algorithms in SU-SC systems when $\NRFt=\NRFr=N_s=4$ and $N_c=30$.\label{fig1}
		}
	\end{figure}
	As a great number of previous efforts have been spent on point-to-point systems, it is intriguing to test the performance of the proposed algorithm by comparing with existing works as benchmarks.
	The OMP algorithm proposed in \cite{6717211} has been widely used as a low-complexity algorithm with the analog precoder selected from a predefined set. The MO-AltMin algorithm was then proposed in \cite{7397861} to improve the performance of the OMP algorithm, yet with high computational complexity of performing the manifold optimization. 
	Both of these algorithms are applied with the SPS implementation. 
	Fig. \ref{fig1} shows that the proposed FPS-AltMin algorithm achieves the highest spectral efficiency with the simulation time comparable to the OMP algorithm. The performance gain is mainly attributed to the proposed FPS implementation, where the unit modulus constraints in the SPS implementation are relaxed. Furthermore, the proposed algorithm leads to an effective design of the dynamic switch network, and provides a better approximation of the fully digital precoder than existing algorithms.
	%Furthermore, the proposed FPS-AltMin algorithm is shown to be effective as it outperforms the MO-AltMin algorithm with a high spectral efficiency in the SPS implementation.
	
	\subsection{Multiuser Multicarrier (MU-MC) Systems}
	\begin{figure}
		\centering
		\includegraphics[height=5.5cm]{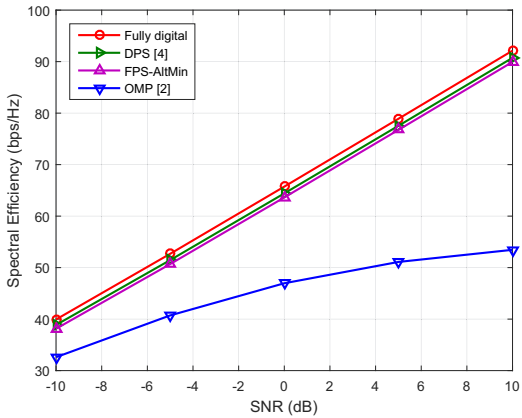}
		\caption{Spectral efficiency achieved by different hybrid precoding algorithms in MU-MC systems when $\NRFt=8$, $\NRFr=N_s=2$, and $N_c=30$.\label{fig2}
		}
	\end{figure}
	In \cite{asilomar}, the DPS implementation was proposed for MU-MC systems to approach the performance of the fully digital precoder by sacrificing the hardware complexity of employing a large number of phase shifters, i.e., $2\Nt\NRFt$ phase shifters. As shown in Fig. \ref{fig2}, the proposed FPS-AltMin algorithm only entails little performance loss compared to the DPS implementation when only 30 fixed phase shifters are adopted. On the other hand, it enjoys significant improvement in terms of spectral efficiency compared to the OMP algorithm. This result demonstrates the effectiveness of both the newly proposed implementation and algorithm. In addition, it indicates that the number of phase shifters can be sharply reduced even if the analog precoder is shared by all the subcarriers and users in MU-MC systems.
	
	\subsection{How Many Phase Shifters Are Needed?}
\begin{figure}
		\centering
		\includegraphics[height=5.5cm]{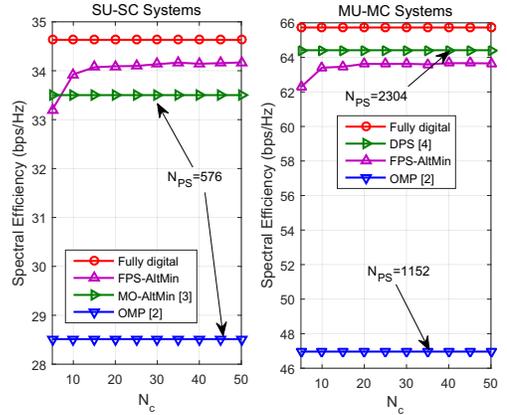}
		\caption{Spectral efficiency achieved by different hybrid precoding algorithms in mm-wave MIMO systems given $\mathrm{SNR}=0$ dB.\label{fig3}
		}
	\end{figure}

	Fig. \ref{fig3} plots the spectral efficiency achieved with different numbers of fixed phase shifters, i.e., $N_c$. The simulation parameters are the same as those in Figs. \ref{fig1} and \ref{fig2} for SU-SC and MU-MC systems, respectively. Fig. \ref{fig3} shows that in SU-SC systems 15 phase shifters are enough for achieving a satisfactory performance as the spectral efficiency almost saturates when we further increase the number of fixed phase shifters. By contrast, 576 phase shifters are needed in the SPS implementation. Moreover, the OMP algorithm achieves a lower spectral efficiency and the MO-AltMin algorithm suffers from the high computational complexity. A similar phenomenon is found in MU-MC systems, i.e., around 10 fixed phase shifters are sufficient, which has not been revealed in existing works. Although the performance of the DPS implementation slightly outperforms the proposed FPS-AltMin algorithm, it employs 200 times more phase shifters. This illustrates that the proposed FPS implementation is much more cost-effective than existing hybrid precoder implementations, and with satisfactory performance.
	
	\section{Conclusions}
	In this paper, we proposed a cost-effective hybrid precoder implementation with a small number of fixed phase shifters. To enhance the performance, a dynamic switch network was adopted, for which a low-complexity AltMin algorithm was developed. The proposed implementation is able to approach the performance of the fully digital precoder, remarkably, with small numbers of RF chains and phase shifters. Thus, this proposal stands out as a promising candidate for hybrid precoders for 5G mm-wave systems.
	
	\appendices
	\section{Proof of Proposition 1}
	\newcounter{TempEqCnt}                         			% 创建临时变量TempEqCnt
	\setcounter{TempEqCnt}{\value{equation}} 			% 将当前公式序号 赋给TempEqCnt
	\setcounter{equation}{\value{longequ}}          	% 当前公式序号变为x，x等于长公式应有的序号减1.
	\begin{figure*}
		\begin{equation}\label{longeq}
		\begin{split}
		&\relphantom{=}f(\alpha)=\left\Vert\mathbf{\tilde{x}}-\mathbf{\alpha s}\right\Vert_2^2\\
		&=\begin{dcases}
		\sum_{j=1}^i(\tilde{x}_j-\alpha)^2+\sum_{j=i+1}^{n}\tilde{x}_j^2&\alpha<0\text{ and }\frac{\alpha}{2}\in\mathcal{R}_i\\
		\sum_{j=1}^i \tilde{x}_j^2+\sum_{j=i+1}^{n}(\tilde{x}_j-\alpha)^2&\alpha>0\text{ and }\frac{\alpha}{2}\in\mathcal{R}_i\\
		\end{dcases}=\begin{dcases}
		i\alpha^2-2\sum_{j=1}^i \tilde{x}_j\alpha+\sum_{j=1}^{n}\tilde{x}_j^2&\alpha<0\text{ and }\alpha\in[2\tilde{x}_i,2\tilde{x}_{i+1}]\\
		(n-i)\alpha^2-2\sum_{j=i+1}^{n}\tilde{x}_j\alpha+\sum_{j=1}^{n}\tilde{x}_j^2&\alpha>0\text{ and }\alpha\in[2\tilde{x}_i,2\tilde{x}_{i+1}]\\
		\end{dcases}
		\end{split}
		\end{equation}
		\hrule
	\end{figure*}
	\setcounter{equation}{\value{TempEqCnt}} 		% 把TempEqCnt中存的公式序号赋回给当前公式序号
	Note that each entry in the switch matrix $\mathbf{S}$ is either $0$ or $1$, and we discover that they can be optimally determined individually once $\alpha$ is given. In particular, to minimize the objective function, $s_{m,n}$ should take value $1$ if the corresponding $(m,n)$-th entry in the matrix $\Re\left(\Fopt\FDD^H\mathbf{C}^H\right)$ is closer to $\alpha$ than $0$ in the Euclidean space, and take value $0$ otherwise, as given in \eqref{eq15}.
	
	The remaining problem is to choose an optimal $\alpha$ that minimizes the objective function.
	Since $\mathbf{S}\in\mathcal{B}$ is an element wise constraint, to simplify the notations, it is equivalent to consider the vectorization version of \eqref{eq13}, given by
	\begin{equation}
	\begin{aligned}
	&\underset{\alpha,\mathbf{s}}{\mathrm{minimize}} && \left\Vert\mathbf{x}-\alpha\mathbf{s}\right\Vert_2^2\\
	&\mathrm{subject\thinspace to}&&
	\mathbf{s}\in\{0,1\}^n,
	\end{aligned}
	\end{equation}
	where $n=\Nt N_c\NRFt$, $\mathbf{x}\triangleq\mathrm{vec}\left\{\Re\left(\Fopt\FDD^H\mathbf{C}^H\right)\right\}$, and $\mathbf{s}=[s_1,s_2,\cdots,s_n]\triangleq\mathrm{vec}\left\{\alpha\mathbf{S}\right\}$.
	
	First, we sort the entries of $\mathbf{x}$ in the ascending order as $\mathbf{\tilde{x}}=[\tilde{x}_1,\tilde{x}_2,\cdots,\tilde{x}_n]$, where $\tilde{x}_1\le\tilde{x}_2\le\cdots\le\tilde{x}_n$. 
	Then all the entries split the real line into $n+1$ intervals $\{\mathcal{I}_i\}_{i=0}^n$, where $\mathcal{I}_i\triangleq[\tilde{x}_i,\tilde{x}_{i+1}]$. Furthermore, we can obtain some insights from \eqref{eq15} to optimize $\alpha$. Specifically, if $\frac{\alpha}{2}$ falls into a certain interval $\mathcal{I}_i$, the corresponding optimal $\mathbf{s}$ can be determined as
	\begin{equation}
	\{s_k\}_{k=1}^{i-1}=\begin{cases}
	0&\alpha>0\\
	1&\alpha<0,
	\end{cases}\quad
	\{s_k\}_{k=i}^n=\begin{cases}
	1&\alpha>0\\
	0&\alpha<0.
	\end{cases}
	\end{equation}
	\setcounter{equation}{\value{longequ}} 			% 当前公式序号变为y，y等于长公式的序号.
	\addtocounter{equation}{1}
	Therefore, the objective function in \eqref{eq13} can be rewritten as \eqref{longeq} at the top of this page.
	Note that within each interval $\mathcal{R}_i=[2\tilde{x}_i,2\tilde{x}_{i+1}]$, the objective function is a quadratic function in terms of $\alpha$, and hence it is easy to give the optimal solution for $\alpha$ in Proposition 1.
	
	\section{Proof of Lemma \ref{lem1}}
	We prove Lemma 1 by contradictory. Since in each interval $\mathcal{R}_i$ the objective function is a quadratic function of $\alpha$. The optimal $\alpha^\star$ can only be obtained at the two endpoints of $\mathcal{R}_i$ or at the axis of symmetry if the objective is not monotonic in $\mathcal{R}_i$. When $\alpha<0$, the axis of symmetry of the quadratic function is given by
	\begin{equation}
	{\bar{x}}_i=\frac{\sum_{j=1}^i\tilde{x}_j}{i},
	\end{equation}
	which is the mean value of the first $i$ entries in $\mathbf{\tilde{x}}$. 
	
	A hypothesis is firstly made that a certain endpoint $\tilde{x}_i$ is the optimal solution to $\alpha$. It means that the axis of symmetry of the objective function in $\mathcal{R}_{i-1}$ is on the right hand side of $\tilde{x}_i$, and the axis of symmetry of the objective function in $\mathcal{R}_i$ is on the left hand side of $\tilde{x}_i$, i.e.,
	\begin{equation}\label{eq27}
	{\bar{x}}_i<\tilde{x}_i<{\bar{x}}_{i-1}.
	\end{equation}
	Note that the entries in $\mathbf{\tilde{x}}$ are ordered in the ascending order. Hence, ${\bar{x}}_i$, as the mean value of the first $i$ entries in $\mathbf{\tilde{x}}$, is an increasing function with respect to $i$, i.e., ${\bar{x}}_i\ge{\bar{x}}_{i-1}$, which is contradictory with \eqref{eq27} and completes the proof for $\alpha<0$. The scenario of $\alpha>0$ can be similarly proved.
	
	\bibliographystyle{IEEEtran}
	\bibliography{bare_conf}
\end{document}